\documentclass{ws-procs9x6}

\usepackage{cite}

\newcommand{\epem}   {\ensuremath{\mathrm{e^+e^-}}}
\newcommand{\sigtot} {\ensuremath{\sigma_{\mathrm{tot}}}}

\newcommand{\sigl}   {\ensuremath{\sigma_{\mathrm{L}}}}
\newcommand{\sigt}   {\ensuremath{\sigma_{\mathrm{T}}}}
\newcommand{\sigch}  {\ensuremath{\sigma_{\mathrm{ch}}}}
\newcommand{\ddel}   {\ensuremath{\mathrm{d}}}
\newcommand{\as}     {\ensuremath{\alpha_{\mathrm{S}}}}
\newcommand{\oaa}    {\ensuremath{\mathcal{O}(\as^2)}}
\newcommand{\qqbar}  {\ensuremath{\mathrm{q\overline{q}}}}
\newcommand{\ash}    {\ensuremath{\hat{\alpha}_{\mathrm{S}}}}
\newcommand{\mz}     {\ensuremath{M_{\mathrm{Z^0}}}}
\newcommand{\asmz}   {\ensuremath{\alpha_s(M_{\mathrm{Z^0}})}}
\newcommand{\azero}  {\ensuremath{\alpha_{\mathrm{0}}}}
\newcommand{\xinull} {\ensuremath{\xi_{\mathrm{0}}}}
\newcommand{\lmqcd}  {\ensuremath{\Lambda_{\mathrm{QCD}}}}

\begin{document}

\title{ Studies of Fragmentation Functions using \epem\ Annihilation
Data from PETRA and LEP }

\author{S. KLUTH}

\address{ Max-Planck-Institut f\"ur Physik, \\
F\"ohringer Ring 6, \\ 
D-80805 M\"unchen, Germany, \\ 
E-mail: skluth@mppmu.mpg.de }

\maketitle

\abstracts{ We present studies of the angular and momentum distributions of
charged particles in hadronic final states of \epem\ annihilation.
Some results are derived from reanalysis of data of the JADE experiment
operating at the PETRA \epem\ collider at DESY from 1979 to
1986~\cite{naroska87} while other data are from
OPAL~\cite{OPALPR362}. }

\section{ Longitudinal Cross Section }

The results shown in this section are published
in~\cite{blumenstengel01a}.  The distribution of the angles $\Theta$
between the charged hadrons and the $\mathrm{e^-}$ beam direction is
predicted to have the following form~\cite{nason94}:
\begin{equation}
\frac{1}{\sigtot}\frac{\ddel\sigch}{\ddel(q\cdot\cos\Theta)} =
\frac{3}{8}(1+\cos^2\Theta)\frac{\sigt}{\sigtot} + 
\frac{3}{4}(\sin^2\Theta)\frac{\sigl}{\sigtot}\;\;,
\end{equation}
where \sigtot\ is the total cross section for hadron production,
\sigch\ is the cross section for charged hadron production, $q$ is the
electric charge of the hadron and \sigt\ and \sigl\ are the transverse
and longitudinal cross sections.  The cross sections \sigt\ and \sigl\
refer to the polarisations of the intermediate electroweak gauge
bosons of the interaction where \sigt\ practically dominates above
quark production thresholds.  A significant contribution to \sigl\
comes from gluon radiation in the
\qqbar\ final state and can be predicted in perturbative QCD in
\oaa~\cite{rijken96} to be
$(\sigl/\sigtot)_{\mathrm{PT}}= 2\ash + 33.78\ash^2$
with $\ash=\as/(2\pi)$. 

\begin{figure}[t]
\includegraphics[width=0.9\textwidth]{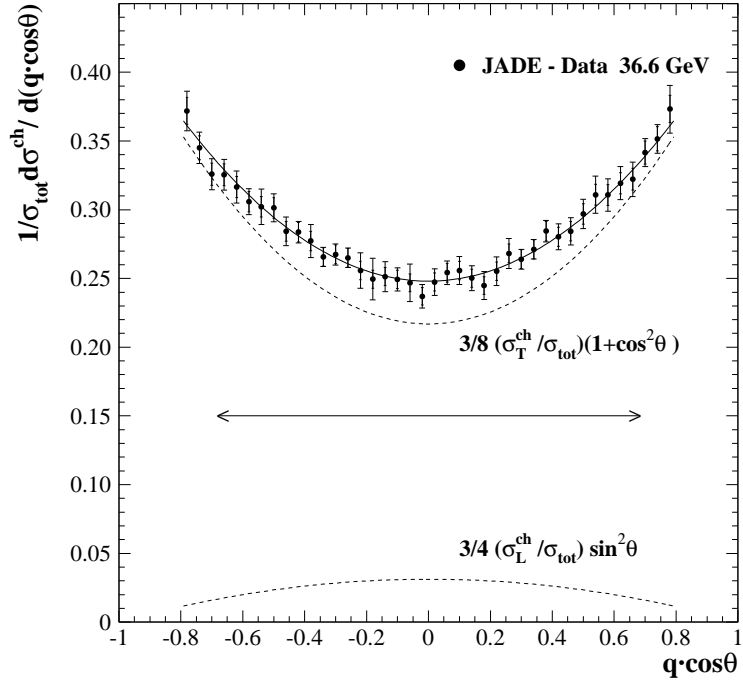} 
\caption[bla]{ Distribution of production angles of charged hadrons
w.r.t. the $\mathrm{e^-}$ beam direction measured with JADE
data~\cite{blumenstengel01a}. } 
\label{fig_sigl}
\end{figure}

The result of the analysis of JADE data taken at $\sqrt{s}=35$ and
44~GeV is shown together with the fit curves in figure~\ref{fig_sigl}.
The fit result is $\sigl/\sigtot=0.067\pm0.013$ corresponding to
$\as(36.6 \mathrm{GeV})=0.150\pm0.025$ or $\asmz=0.127\pm0.018$.

Possible effects of hadronisation have not been considered in this
analysis.  In~\cite{beneke97} hadronisation effects are predicted to
scale as a $1/Q$ power correction where the relative size of the
effect is given by a universal parameter \azero.  A fit of this
prediction to data for \sigl\ from OPAL, DELPHI and this analysis
yields $\asmz=0.126\pm0.025$ and $\azero(2 \mathrm{GeV})=0.3\pm0.3$.
The uncertainties of the present data do not allow to draw conclusions
about the power correction.

\section{ Momentum Spectra }

\begin{figure}[t]
\begin{tabular}{cc}
\includegraphics[width=0.45\textwidth]{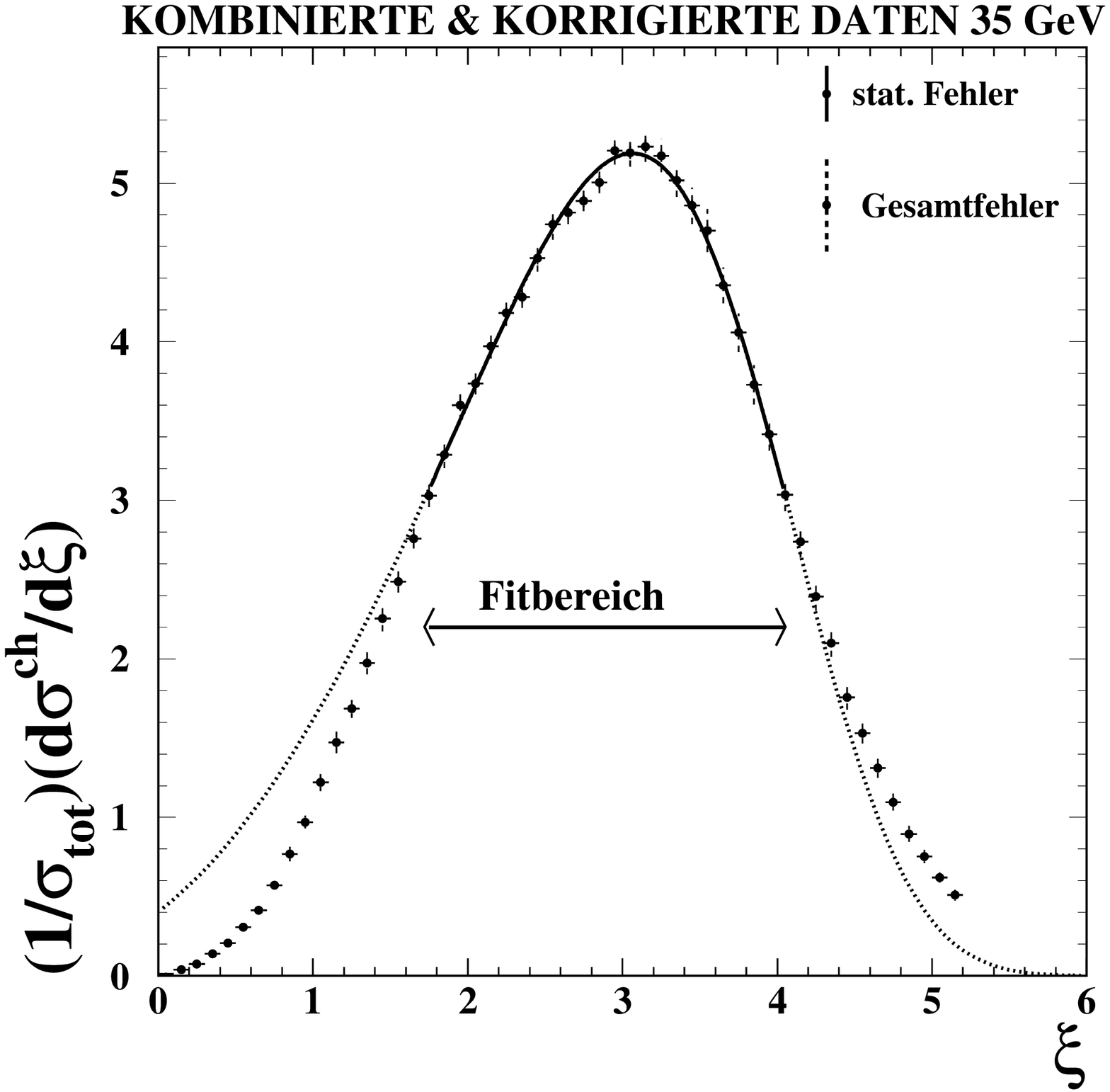} &
\includegraphics[width=0.45\textwidth]{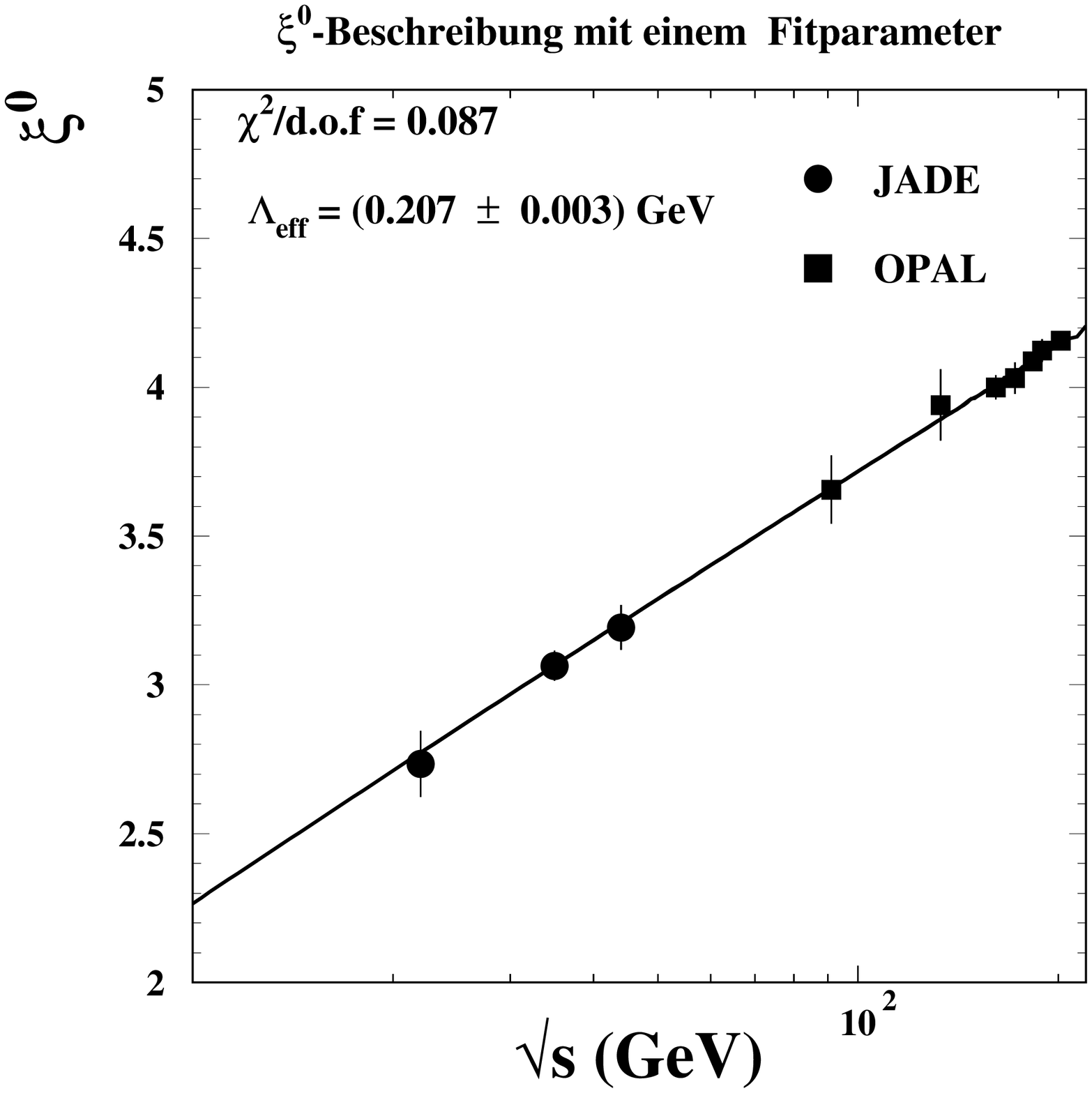} 
\end{tabular}
\caption[bla]{ Distribution of $\xi$ measured by JADE at 35 GeV (left)
and peak positions \xinull\ as function of $\sqrt{s}$
(preliminary)~\cite{blumenstengelphd}. } 
\label{fig_xsi}
\end{figure}

The study of momentum spectra of charged particles in hadronic final
states of \epem\ annihilation allows tests NLLA QCD predictions
together with Local Parton Hadron Duality (LPHD)~\cite{Fong91}.  For
charged particles with momentum fractions $x=2p/\sqrt{s}$, where $p$
is their momentum, the distribution of $\xi=\ln(1/x)$ is studied; here
we show results using data from JADE and
OPAL~\cite{blumenstengelphd,OPALPR362}.  Figure~\ref{fig_xsi} (left)
shows the $\xi$ distribution measured by JADE (preliminary) at
$\sqrt{s}=35$~GeV while figure~\ref{fig_xsi} (right) shows the
evolution of the peak position \xinull\ with $\sqrt{s}$ using JADE
(preliminary) and OPAL data.  The evolution of the peak position
\xinull\ is seen to be well described by the NLLA prediction.

The NLLA QCD prediction is found to describe the data at 22, 35 and
44~GeV well around the peak region where the predictions are expected
to be valid.  A more detailed study of LPHD was done by
OPAL~\cite{OPALPR362} by fitting MLLA predictions~\cite{khoze97} which
also predict the normalisation.  A so-called hadronisation constant
$K^h$ gives the number of hadrons produced per parton and is seen to
be constant for $\sqrt{s}>\mz$.

In~\cite{blumenstengelphd} possible effects of finite quark masses on
the $\xi$-distribution are studied by introducing effective parameters
$\lmqcd^f$ with $f=\mathrm{uds,b}$ or c to absorb mass effects.
Using OPAL data for $\xi$ measured for uds, c or b quarks as
constraints the effects of the b quark mass w.r.t. light quarks are
indicated as ca. 25\% in the values of $\lmqcd^f$.


\end{document}